# What is Time?
## A New Mathematico- Physical and Information Theoretic Approach

**G.SURYAN, Physics Dept., Indian Institute of Science, Bangalore 560012, India.**

Introductory remarks

This paper is the first one of two related papers.

The second paper entitled "***Estimation of the Gravitational constant $G_N$ from CMBR data and the Hubble's parameter***" follows. In addition, relevant notes and clarification remarks, pertinent references & select Bibliography with quotes from various authors are appended to each.
It may also be noted that the two main papers are written in a form that can be easily understood by an intelligent person and can be read independently. The existing superstructure of physics is not disturbed. The paradigm shifts and their role have been brought out in the form of Block diagrams and a few figures.

The first paper follows

## What is Time?
## A New Mathematico-Physical and Information Theoretic Approach

Examination of the available hard core information to firm up the process of unification of quantum and gravitational physics leads to the conclusion that for achieving this synthesis, major paradigm shifts are needed as also the answering of 'What is Time?' The object of this submission is to point out the means of achieving such a grand synthesis. In the light of overwhelming literature[1-11] accumulated over many millennia through the deep philosophical thoughts of saints, savants and scientists of various persuasions, such claims may be thought of as foolhardy and brushed aside. It is hoped that some readers would have the objectivity and curiosity to read further. Currently the main pillars supporting the edifice of physics are : i)The geometrical concepts of space-time-gravitation ii) The dynamic concepts involving quantum of action iii) Statistical thermodynamic concepts, heat and entropy iv) Mathematical concepts, tools and techniques serving both as a grand plan and the means of calculation and last but not least v)Controlled observation, pertinent experimentation as the final arbiter. In making major changes the author is following Dirac's dictum "….make changes without sacrificing the existing superstructure"[12] Box (1) . There have been many attempts at geometrising the universe of which the geometrodynamics of Wheeler is a good example. To quote Feynman , "Although geometrodynamics as developed by J.A. Wheeler & co-workers have not yielded any quantitative results, it contains seeds of bold imagination which may yet lead to spectacular successes in our understanding of physics".

For this a geometrical interpretation of the Planck's quantum of action and its primacy as well as paradigm shifts as shown in Boxes 1, 2 & 3 have been introduced and it is shown that Time can be understood as a parameter **within the confines of a finite growing universe.**

The approach involving the "Ekon" and "Chala-Chala" concepts clarifies the essentially asymmetry of the nature of time and what is gravitation by demonstrating by numerical estimates the gravitational constant and also the possible structure of the most stable particles electron and proton . The relationship between statistical thermodynamics and the arrow of Time is also brought out.
The requisite mathematical apparatus and its relations to the existing super structure of physics has been brought about by utilizing the existing Hamiltonian quaternion formalism with the scalar part vastly large. The normal equation of the Einsteins second order differential equation is recovered at sufficiently large scales.

## Major paradigm shifts introduced are as follows :
**i)Upper bound :** Provide for only a finite number of finite entities implying coarse graining. In the words of Ashtekar [13] " the loss of continuum is such a dramatic effect that it is bound to have a deep impact on science". Why introduce infinities in the first place and later subtract them? ii) **Lower bound :** Avoid Zeros: they are even more dangerous than infinities. They are introduced innocuously. Firstly as a place value really denoting multiplication or division. It would be necessary to specify what is that which is zero. iii) Reciprocal relationship : An upper bound for Space is linked to a lower bound for Momentum and vice-versa. Emphasis is on *position and momentum* jointly rather than energy and time. iv)Major role of repulsive Interactions. v)New mathematics is needed to link conventional relativity mechanics theory to quantum mechanics. All these above points are being incorporated as follows.

**The Arena :** Though the much used word 'space' in its general sense may serve, it is better to use altogether different word ; hence "The Arena". Appropriate names and properties for the new fundamental physical entities and their assemblies follow.

**Fundamental building blocks** of the universe are named as EKON, from the Sanskrit word EKOM meaning single. The only information EKONS have is about their own identity and assigned value of $[ \frac{1}{2} *(h/2\pi )]^3$ *having a dual character of both rest and motion*.
Assembly of EKONS is given the name CHALACHALA, *CHALA* meaning motion and *ACHALA* meaning rest and its finite extension as collection of a large finite integral number of EKONS each of value $[ \frac{1}{2} *(h/2\pi )]^3$ all jostling about. Two or more EKONS can join only end on and may even loop and knot by joining with their tails. Because of the mutually reciprocal character of position **r** & momentum **k**, **_confinement in one leads to an extension of the other but keeping their product constant_**.

This is an "affinity" which gives the concept of a neighborhood and one can formulate a Hausedorf 'space' and measures of distance, area and volume.

Consider a simple act of picking up an object and placing it at a higher level support. Note that all the energy supplied by various sources in our body have been fully converted to the potential energy of the object and can be reconverted to any other form of energy like kinetic, thermal, chemical and so on. What are the properties required of the entities or medium for this to happen? No matter the energy is tiny weebit or huge, the mutual transformation from kinetic to potential must be taking place at all scales. What is the smallest scale at which this takes place so that larger effects on larger scales may be viewed as repeated acts of the small? Such smallest entities have been named as "EKONS". They are the embodiment of both rest and motion and their assembly constitute the space called "CHALACHALA". Motion is of two kinds one due to the linear motion and the other due to rotation and "spin".

The substratum 'CHALACHALA' introduced earlier ***should not be mistaken for the normal phase space of current physics literature***.
In the present approach the growth in the number of basic entities is taken as a measure of Time. The net number of the fundamental entities created up to any point of Time is denoted by $N_W$ (W for Wheeler).
The fluctuations in this number $[ (\Delta N_W/N_W)^2 ]_{av}$ is a measure of uncertainty in 'Time" which of course will be very small because $N_W$ is very large making it appear that the parameter can be taken as almost continuous, thus preserving the common sense of the word Time. Growth in numbers preserves the unidirectional Time.
People conditioned by conventional thinking have no difficulty in imagining an upper bound for energy and momentum. However it requires a mental effort to imagine an upper bound for space. There is no need to bring in infinite space except as an aid to use the powerful techniques of mathematical analysis through differential and integral calculus which brings in their own "infinities" and the task of using them with caution.

Presently subtracting infinities has been carried out to the extent of being an 'art form'. Equations are written in a global variational form whereas if one looks carefully they all mean some local minima.
Physical systems do move on to states of other minima far removed from an initial local minimum. Large changes can result from small inputs. One needs to have a mental approach to accommodate it. In some sense the ideas put forth here is in consonance with ideas propounded by Ilya Prigogine [14].
Linear motion can be annulled by comoving. Spin motion is intrinsic though visualized as rotation, their mathematical (geometrical/analytical) representation is to be defined carefully. Note that such spin can be canceled only by an equal and opposite spin, geometrically visualized as a "twist' and the half twist is the most durable.

**Key points regarding the EKONS may be listed as follows :**
i)An ordered pair of numbers ii) built in duality, iii) Visualized as a oriented spiral element, half Twist iv) Like half-twists can join together or v) opposite half -twists can just disappear leading to random creation and annihilation vi) Net creation identified as TIME parameter vii) each assembly of EKONS is its own clock (they don't have a quartz watch on their wrists or measuring rods, protractors, book of fundamental constants!) viii) assemblies of such EKONS are visualized as an "affine" space ix) The essence of uncertainty lies in being a right or left twist.

Homogeneity and isotropy of CHALACHALA would be dependent upon the scale of observation' and such scales are themselves assembly of EKONS which brings up the importance and need for an information theoretic approach.

How the basic difficulty of reconciling the Heisenberg's quantum theory with Einstein's gravitation theory can be overcome is shown in the accompanying Box(4) assigning the primary role for the Planck's Quantum of action followed by Heisenberg's uncertainty relation.

This is a major paradigm shift. Intelligent lay people would certainly wonder about physicists and how funny they are, who swear by symmetries of all kinds and then break them! This leads to another paradigm shift i.e ***not to use any symmetries (and break them) but to show how symmetries can arise as a consequence of the appropriate length, momentum, time, scale of observation and theoretical modeling adopted.***
Here it is appropriate to refer to the beautiful and highly relevant experiments by Gollub[15] et.al on surface waves in fluids which demonstrate how symmetries can arise as a result of averaging over some time scale. In the current literature of theoretical physics, symmetries of various kinds are invoked such that with each a conservation principle is associated and the alluring powerful group theoretical techniques can be used. Rich mathematical apparatus has evolved around these groups like semi groups, Lie groups and so on. One is led to believe that the physical systems have the information capacity to make sure that the model systems behave in the right manner. The continuous, infinitely sub divisible and smooth space -time brings in lot of powerful mathematics particularly the differential and integral calculus along with their infinities. From the information theoretic view the situation is just not affordable. Finite and coarse graininess has to be accepted.

**Assumptions and procedures for further estimates follow:**
i)All EKONS are alike with a "volume" $[1/2 * (h/2\pi)]^3$ each and do not overlap thus constituting CHALACHALA. ii) The number $N_W$ is increasing and is identified with The increase parameter we call Time iii)EKONS obey a form of statistics combining Bose - Einstein and Fermi -Dirac statistics, approximated to Fermi-Dirac statistics for purposes of numerical estimates. iv)An effective temperature $T_m$ is associated with this. For cold dark matter, high degeneracy is expected v)The width of the derivative function at the effective Fermi level, gives the effective Temperature.

vi) EKONS by themselves have no mass and it is postulated that but gain inertial mass by invoking the strong Mach's principle; the prefactor in the Fermi Dirac distribution will give an *effective mass*' $m_\varepsilon$ vii)The Fermi level of the universe therefore gives the age of the universe and also the "Chemical potential" viii)The width gives a measure of the expansion of the universe and is related to Hubble constant. ix) $N_\varepsilon$ (the number density) x $m\varepsilon$ gives the Mass density term $\Lambda$ introduced by Einstein x)The 'volume' of the universe is $N_w*[1/2 (h/2\pi)]^3$. Note that this "volume" refers to the joint, momentum/space, combine.

**Interpretation of (h) as an angle (Ratio)**
In understanding the ubiquitous ½ , the concept of Twist is useful. Twist has a direction associated with it as left or right. A half twist combined with another half twist becomes one. Alternately two opposite half twists can cancel each other and that information is completely lost. Hence another paradigm shift and need for information theoretic approach. Each system big or small is its own clock. Irreversibility and homogenization of phase space are the consequences of : i) Very large numbers ii) thermal equilibrium iii) quasi - steady state iv)quasi - continuous time parameter v) small changes can lead to large consequences depending upon the scale of Time, distance vi) values of 'fundamental constants' are no longer independent of time, an idea put forth by Dirac in a most forceful and prescient way Box (1) .

It is necessary to remember that certain invariance principles have to be invoked to enable a mathematical approach to the following physical ideas as follows :
  A. Invariance with respect to choice of origin in space. This is easily understood for nature in its intrinsic property is no different when it is viewed from different positions.
  B. The physical systems do not alter because of the uniform movement of the observer and this goes by the name of Galilean principle of relativity.
  *C.* The next one was introduced by Einstein as the principle of equivalence : <u>uniform acceleration should also not matter</u> . To put this idea in a mathematical form Einstein had to rely on Reimanian geometry. In Quantum dynamics it is usual to treat position and momentum are related as x and $\partial /\partial x$ ; that is momentum as derivative, *and choice of origin for momentum should not matter!* Einstein met this requirement by using four dimensional Reimanian geometry suitably adopted to accommodate Time along with the square of velocity of light and a minus sign, as another dimension. This requires extra terms to be added to the ordinary derivative. This is called a 'connection' which makes the space "curved" and is identified with universal gravitation and a smooth manifold is assumed having infinitely many derivatives, *a luxury we must give up.* In the present approach there is no need to use extra time dimension and it is enough to treat Time as a *parameter*. *Three dimensions are just enough.*

The nature of Time as a parameter is best understood by noting that nature provides its own clocks. Two important requirements for a clock are (i) some phenomenon which repeats itself (ii) some means of storing the number of repeats. Best geometrical representation is in the form of an Archemedian spiral growing outwards or shrinking inwards, $r = k\theta\, e^{i\theta}$. The growth or decay factor is absorbed in the exponential using complex numbers. It is to be noted that a complex number is just an ordered pair of numbers.

Quasi periodic motion can either be "growing spirally" or contracting and would enable a count of the periods to be kept. In addition there has to be a means of ascertaining the direction of rotation providing that the spiral may grow or contract. In all four numbers are required. In an assembly of several non interacting entities the resultant mathematical representation would be a simple product of the amplitudes and sum of the exponents which may not all be in synchronism with one another.

If the repeats of the positions are indistinguishable from a previous position the system would be classed as cyclic and it would be possible to mention about time only as modulo of a characteristic period the integral part being lost.

The longest growth/decay would represent the growth or decay of the entire system. When the individual items can be factored in to identical parts the amplitude factor will represent the total number and the phase factor will give the small changes.

The ultimate uncertainty would be one entity. Because of the large numbers involved a Time parameter and a corresponding position and momentum variables can be taken as smooth . In Einstein's approach the functions $g_{\mu\nu}$ and the $x^{\mu}\, x^{\nu}$ are considered smooth and with as many derivatives as required . The new mathematics as given below, the Einstienian formulation is being modified to be suitable for coarse grain applications.

## What is the New Mathematics ?

In order to link the present ideas with the well developed formulation of general relativity it is necessary to have new mathematics. It is common practice to write expressions like $1+ a_1 x + a_2 x^2 + a_3 x^3 + \ldots\ldots$ as applied to physics problem say involving length x , the meaning of $x^2$ is area, $x^3$ is volume and so on. How can one add a length with an area or volume ?

It is normally assumed that such issues are covered by the coefficients a's. *They must have precisely the reciprocal of the dimensions corresponding to the powers of x* . This stipulation applies whatever the exponent of the terms be. If there are more than one variable say like $x_1$ , $x_2$ then the expression would be like $g_{\mu\nu}\, x^{\mu}\, x^{\nu}$ (with summation convention used in tensor calculus and set of $g_{\mu\nu}$ 's must have inverse dimensions to those of x's).

In the differential form the expression is used effectively to make sure that the distance between two neighboring points in a space is invariant with respect to the choice of the reference axis x which is arbitrary and generally chosen to conform to the perceived or explicitly desired symmetries for the calculation on hand.

**Applicability of the Mach's principle in its strong form :** This principle is best understood with reference to the accompanying cartoon depicted in Box (5). It is quite clear that whatever little change takes place will have a reaction and this has to take place instantaneously that is the rest of the universe has to react. For this it is necessary to refer to Mösbauer phenomenon where the reaction to the emission of radio-active particle can be shared by the entire system and a similar thing has to happen at the cosmological level also.

**Importance of Repulsive interaction :** The only information that a single EKON has is its own identity. While two EKONS can join end on they will not overlap. When they approach each other sideways there is no chance of any attraction; they just keep their identity. Thus when assemblies of EKONS come near each other they have to move on to larger momentum values. Each will move away from the other, that is they repel each other and this is a cosmic repulsion.

The idea of the fundamental repulsive interaction is not an adhoc assumption. It is rooted in observation. An elegant example is from magnetic lines of force. Each of these lines of forces repel its neighbor. In fact this has resulted the idea of flux quantisation. It may be surmised that this type of repulsion which allows the expansion of the universe; and which can take place at all scales. The world can be finite but expand due to random net creation of such EKONS and their assemblies takes place. Thus expansion of the universe discovered by Hubble is easily understood. Using the known Hubble parameter one can estimate the age of the universe and also estimate the net number of EKONS generated so far. Probably generation is a bulk process rather than on the edge of an expanding universe.

**Recovering the Energy concept and its relation to Time**
Having recognized the time concept as a parameter, it is necessary to rightfully recall it to its role in order to make <u>contact with existing superstructure</u> of physics in a self consistent manner and also to <u>have a workable scheme</u> to get numerical results for comparison with observation and experiment. One of the most important applications of the EKON concept introduced is understanding the phenomenon of optical interference say from a grating. In most of the discussions on this topic the general assumptions is made that such gratings, mirror and other optical equipment used is classical and therefore it has no part to play. *This assumption is not valid*. The new approach is that <u>even the so called classical objects are composed of EKONS and their assemblies with the built in dual character</u>. In consequence, the EKONS constituting the gratings there would have a corresponding extension in the momentum portion of the entities . These in turn would guide the impinging stream of EKONS constituting the input optical radiation.

The first major question would be "If Time has been disengaged from Energy and the mass is due to the rest of the universe what happens to the relation $E = mc^2$ which has been amply verified not only in physics but also in war as a nuclear weapon with disastrous consequences ?". These questions form is subject matter of a very promising study by the author . However the author wishes to publish it separately after ascertaining the views of erudite conversant with the subject.

In a penetrating analysis of the concept and interpretation of The Arrow of Time in a book by Huw Price[16] entitled "Times Arrow and Archimedes' Point", he has questioned as to how entropy is low in the first place. The present approach the Ekon and their assembly CHALACHALA are essentially cold just as an lifted object, a rotating flywheel, a charged super conducting magnet are all essentially 'zero' temperature and low entropy. The randomness comes about essentially as a consequence of random net generation.

In this connection reference may be made to a critical but generally appreciative essay review on Huw Prices's book by HR Brown[17] is also appropriate. The main question raised as to how the unidirection of time is ensured and also conforming to thermodynamic arrow has also been brought out in the companion communication under the title **Estimation of Gravitational Constant 'G$_N$' from CMBR data and Hubble's parameter**.

**Conclusions and summary :** A scale independent geometrical view of Planck's Constant and Heisenberg's Uncertainty relation and its prime importance has been proposed, leading to the concept of EKON and their generation linked to Time as a parameter than as an additional dimension. Major paradigm shifts have been introduced not dependent on abinitio symmetries, symmetries arise on time, distance or momentum scales of observation. A simple easily understood finite model of the universe in its widest possible range spanning the very small to the very large scales has emerged. A new mathematical apparatus as well as its links to the extensive and deep insights already provided by both experiment and theoretical efforts <u>without compromising the existing super structure of physics</u> has been proposed. Close links between physics and mathematics in the form of built in duality in both has been pointed out. Many of the conundrums of the quantum theory get clarified and easily understandable and opens the way for further verification and deeper understanding. A program for estimating the fundamental 'constants' of nature directly from cosmological data has been initiated. Links to mathematics through group theory and algebra and topology have been indicated. The situation is highly reminiscent of the period 1924-1925 ably summarized by VanVleck[18] in a not so well known report published by NBS."…quantum theory is so alive that it changes almost over night"

**Credits**
Author acknowledges help given by Prof.E C G Sudarshan to coin the word EKON from the Sanskrit word EKOM meaning single and also bringing to my attention Finkelstein's remarks about Twists and their cancellation.
Dr. B R Nagaraj of the Tata Institute of Fundamental Research, Mathematics Center at IISc. campus Bangalore, for drawing my attention to several aspects of modern mathematics particularly in handling step functions and the use of the Gaussian form.

**Acknowledgments**
I acknowledge my indebtedness to the Indian Institute of Science, Raman Research Institute , National Institute of Advance Studies, Bangalore University and other Institutions in India who have given me liberal access to their libraries.

# Additional Notes and comments of relevance

## Notes on time

In the current physics literature 'Time' is used along with position variables as another dimension and in order to balance the dimensions the velocity of light is used. In addition, in order to define a line element in the **r** , **t** space squares of the space part is combined to the squares of the Time part but with a minus sign.
Even though all the terms are space like still it is necessary to resort to the velocity of light as a linking factor between space and Time. Powerful procedures of the tensor calculus enables a formally invariant description and is brought about by using squares of the quantities.

## The role and nature of time as a parameter

It is safe to assume that there is a parameter in the description of most dynamical systems which may be identified with physical time.

Some examples may be in order.
a) Decay and growth of motion : particularly useful for long periods  b) Radioactive decay c) Oscillation/recurrent behavior: Useful for short periods, rotation of pulsars
d) Rotational motion : useful for very long periods, Rotation of earth/sun/stars/
galaxies, each one of these may occur simply or in combination. Therefore it is postulated that every dynamical system has at least one characteristic which can be used as an internal clock. In order to be useful as a clock there must be some means of recording such that the time elapsed may be known and also to enable comparison of clocks. Also there must be overlap of the applicability of each in ranges adjacent to them so that inter-comparison can be made and enable linking the fastest clocks to the slowest. Another necessary feature is the means of dividing larger intervals into 'equal' smaller intervals. This is required for attaining necessary precision, improve upon it to the next level. Once a suitable primary time standard is chosen it would be possible to ensure accuracy.

## Graphical Representation of universal time through the Archemedian Spiral : Polar Plot (Box 6)

While for random events the Archemedian spiral can show jumps anywhere, in actual practice only angles $\pi$ would survive to continue the spiral.
The Polar Plot if viewed as an expanding spiral from inside (anti-clockwise) would correspond to the growing universe. Each sector corresponding to a net generation of one Ekon. On the other hand if the spiral is viewed in clockwise from the outer most limb contracting inwards would correspond to a decrease and roll back of the universe.

Note -1 : Even random events can form the basis for measuring time provided they are sufficiently large in number. For example Radio active decay along with the accumulated daughter products has been used as an internal clock.
Note -2 : 'uniform' motion can be perceived and quantified only if there are better standards with respect to which the motion can be validated. Of course inter comparison of set of clocks is practiced.

## Definition of a clock and Some examples of natural clocks

Clock is a devise which has  i)  A period and counting arrangement
ii)  Something to ensure "equality of periods"

i)Earth's rotation daily about itself / seasons annually about the sun
ii) Sedimentation  iii) Radiant cooling iv) Radioactive decay v) Chemical clocks
vi)Growing plants vii) Heart beats.

## Role of Time -Energy vs Position-Momentum Uncertainty relation
## Notes regarding energy

1. Only energy differences that matter
2. Current practice is to assume that energy of a particle is zero at infinity
    Either energy is positive and tends to zero from above
    **Or Energy is negative and tends to zero from below**
3. A system undergoing a transition form a higher energy state to a lower energy state gives off a  positive energy entity like a photon or other particle and vice versa.
4. Shorter wave lengths are associated with higher momenta and momentum conservation is also used.
5. In a simple two body collision in general it is not possible to satisfy both energy conservation and momentum conservation(including spin \ and this requires a third entity essential just as coarse graining has to be resorted to at small length scales there has to be cut off at high momentum scales  also.
6. Mass is the consequence of the reaction of the entire universe and it is best rely on momentum conservation rather than as energy conservation.
7. In order to make contact with the existing super structure, energy given away is assigned a mass equivalent through the relation $E=mc^2$.

i)In this connection some remarks on the naïve use of $\Delta t \cdot \Delta E \approx h/2$ as fourth uncertainty relation  means that time is also treated as an operator  ii)However, in standard  wave mechanics calculations, time is used as a parameter and found to give numerical results with surprising internal consistency and accuracy, time is just a parameter

iii) It is possible to think of and use an universal time parameter with origin and direction but it is not possible to use a unique energy; only differences of energy are measurable
iv) To preserve Energy conservation is only in principle , short fall is attributed to Dissipation of energy through radiation, frictional forces, collisions etc. is always taking place, in forms which do not allow recovering it.  Also  the quantity of energy depends on the particular path (in thermodynamic sense) it  is cycled, therefore  not a perfect differential and  reciprocal of temperature gets in as a integrating factor  for which there has to be a source and sink at different Temperatures.

**Additional explanatory  information  on EKONS and CHALACHALA**
Transport of energy from one place to another takes place by  transformation kinetic to potential to kinetic……..  In the case of the simplest "sound waves ", the material medium compresses, relaxes, compresses, relaxes.

In  the case of electromagnetic energy, electric field - magnetic field,  electric field  - etc. each of which has its own cycle of potential kinetic components.
In the case of "matter waves" which are supposed to travel essentially in a vacuum how does they move ?  This question is easily answered by noting that EKONS have the character of both space and momentum including angular momentum.  The transport of momentum and angular momentum takes place within the EKON.  One might picture this in the following way.

If the EKON is squeezed latterly by others it would lengthen axially and *vise-versa*. This would correspond to  a change from the kinetic to the potential.  This is going on continuously, no dissipation is involved and there is no natural frequency for this.  The deformation mentioned above does not cause any reaction except that internally the EKON has become either momentum like or space like. As a result the only thing that EKON do is to push away its nearest neighbor.  In conventional physics literature a damping or  growth  factor  would  be  invoked. In the new model now being proposed loss or gain of EKONS would have to be there.  The gain would be by generation and loss by cancellation with an existing EKON of opposite orientation.  Therefore in a system where there is neither growth or diminishing of the numbers of EKONS the system has to be periodic with its own time scale. There would be  loss when EKONS (twists) meet with opposite EKONS with an opposite twist.  Hence the importance of having  a  net  spontaneous  generation  of  EKONS  associated  with  a  unique  time direction.
By movement of EKONS one must understand its own changes from Kinetic to Potential, Potential to Kinetic and so on.  It should not be imagined that EKONS move in an empty space.

They can impart 'motion' to neighbor entities and all wave propagation or motion has to take place with the assistance of other EKONS such that their individual identities  are preserved except for the increase of numbers due to net generation/ identified with Time and the growth of the universe.

In a simple two body collision in general it is not possible to satisfy both energy conservation and momentum conservation (including spin) and this makes third entity essential. For example in Compton effect an x-ray quantum is scattered by an electron and suffers a loss of energy. For this to happen the electron needs to be bound to an atom. A free electron cannot show Compton effect because it is not possible to satisfy conservation of both energy and momentum.

The inverse problem of pair creation by x-ray depends upon both the energy conservation and momentum conservation near a massive nucleus which allows number of parameter to satisfy both energy and momentum conservation.

Presently mass standard is based on a certain quantity of well defined volume of material and weights are compared by precision balances. Thus both length and mass units are involved as also the acceleration due to the gravitation though not the Gravitational constant explicitly. Similarly in the determination of the mechanical equivalent of heat torque are generated by known masses or by electrical means by known electric currents and calibrated electrical resistance which in turn are dependent on Quantum Hall conductance.

A small initial asymmetry towards what we now call 'left' has resulted in the present 'space' dominated universe. A small asymmetry could have resulted in a momentum dominated with time running backward.

## Mathematical Formulation : Lagranian vs Hamiltonian approach

There is much confusion and loose phraseology in the categories of Lagrangians and Hamiltonians. The first one is essentially a classical function of position and velocities and time. The second one is a function of position, momentum and time.

| Lagrangian | Hamiltonian |
|---|---|
| Variables   q. q. t | p, q, t |
| Difference between Kinetic energy and potential energy Function of position & velocity and time | Sum of Kinetic and Potential energy as |
| $L(q.q.t) = T-V$ | $H = T + V = H(p,q,t)$ |
| Minimal principle in the form of variation of an integral Second order derivative Second order differential equation | Differential equations of first order derivatives both ordinary and partial |

Variational principles are put in a form as though the physical systems know the entire Lagrangian or Hamiltonian in advance and are able to find the external paths.

Physical systems just do not have that capacity of information storage and rapidity of computation. In order to put the new ideas into a systematic formulation it was found necessary to evolve a mathematical apparatus at two levels. The first one is in invoking fractional differential and integral calculus. In current (conventional) quantum mechanics the position 'x' and momentum 'p' do not commute and the limit to this is set by the quantum of action ( which is a unit of angular momentum).

In the new approach a situation is envisaged where the position and momentum are treated on an equal footing and it is possible to think of an entity which has _both position and angular momentum characteristics together_ and any one of them can manifest itself more by constraints on the other. The most equitable situation occurs when the representation is in terms "spinors". Half order differential and integral mathematical quantities follow and they are conveniently related to the beta functions. Thus a powerful mathematical apparatus is at hand.

However, the differential form and the limit processes associated implies a continuous space and does not provide for dealing with finite objects like the physical entities EKONS introduced earlier. To handle them it is necessary to think of an 'affine' space in the sense that the angular momentum and position of the ultimate entities are linked to each other in a "reciprocal" relationship.

Assembly of such entities has already been named CHALACHALA. The description (geometry) of such spaces in general can be handled simply and elegantly as follows. The only stipulation made is that we can add only like quantities.

## Some general applications to other discipline like Fluid Mechanics(Turbulence) , Economics, Sociology etc.

An example would make the basic idea clear. Consider the price of cotton on a day to day basis and its relation to stocks on hand. A simple plot would show very large swings even through the hours of one day. What is the sort of "space" that can be considered to predict long term trends. One of the simple methods adopted is to choose a reference year or month as the basis for both the price and stocks and use ratios.
Even then the graph would be highly irregular. In order to smooth the irregularities (fluctuations) it is necessary to use some averaging procedure say over weeks, months or years. The right choice of averaging period would be crucial. An averaging period of one year may be suitable to relate the prices/stocks to say the sunspot cycle, but if the forecast is required for shorter periods of weeks or months corresponding averages must be chosen. Suppose one more variable like rainfall is to be taken, things become pretty complicated.

When Hamilton introduced his quaternion questions were raised as to the funny character of his creation "how could a vector combine with a scalar? It is now possible to answer this query by reinterpreting the i, j, *k* 's as having the precisely the reciprocal "dimensions" of the vectorial components, there by turning them into a scalar.

This aspect has been taken into account in formulating a coarse grained, quasi continuous manifold.  The position adopted by the present author is I) the main scalar part is vastly large increasing numbers  ii)  there is  statistical thermodynamic equilibrium with a characteristic temperature and  entropy.

"Category" consists of all objects of a certain sort (groups, topological spaces etc. and of the mappings (homomorphisms, continuous maps etc. that  exists between the objects (Refer Box 7)

## Fundamental Importance of Planck's  Constant  *h*

Note 1 . *h* was introduced by Plank to link the high frequency behavior (of the black body radiation) and the low frequency behavior each of which had divergent behavior. Thus h appears to be more fundamental than "time".  It should also be noted that in Planck's formula of the "black body" radiation *h* does not appear alone but with  k, Boltzmann's constant  as h/k  which indicates  deep connection with statistical mechanics and thermodynamics.

Basic discoveries have been made involving  h  either as h/e or h/k. In considering the large scale structure of the universe and large time scales, the role of free electric or magnetic or free subnuclear matter like gluons etc. would be small.  The main consideration would be the statistical mechanics of large number of identical  entities  without any charges but with their own  inertial property.

## Earlier attempts at a finite/growing universe and incorporating gravity/thermodynamics gravity and quantum the following need special mention and special remarks from various authors.

**Eddington** :" There  have been many attempts at understanding gravity and integrating with quantum mechanics.  The most notable of these is the attempt by Eddington  in his fundamental theory with emphasis on a finite number of basic entities.
There is only one law of nature - the second law of thermodynamics which recognizes distinction between past and future, one way property of time "vividly recognized by consciousness".

To quote  Eddington  "Fundamental theory" regarding "the necessity of treating *x* & **p** as a joint observable: owing to their non commutation is the principal reason for employing wave rather than distribution function.

Associated with $f(x)$ there is a $F(p)$ in the range $dx$, $dp$

$2\pi h \mid f(x) \mid^2 dx$ and $2\pi h \mid F(p) \mid^2 dp$

$f(x) = (2\pi h)^{-1/2} \int_{-D} \exp(ip(x)/h)\, f(p)\, dp$

$F(p) = (2\pi h)^{1/2} \int \exp(-ip(x)/h)\, f(x)\, dx$

The trouble in using distribution function is that even if $f(p)$ is a regular function of $p$ its complex argument (phase) may be erratic and when there are several "*independent*" particles in the distribution. If for some origin the phase $F(p)$ is constant for all $p$ distribution forms a symmetrical wave packet corresponding to a fully observed particle".

However in the present approach fundamental entities are not independent, they have to respect the independence of their neighbors and this leads to an effective short range repulsion.

**Born** : The attempt by Max Born to write a metric with momentum rather than space-time variables. "A Consequence of an assumption of a finite size of a system in the p-space is the existence of a set of proper functions $\Psi_n(p)$, where the index $n$ refers to proper values of some functions of the space co-ordinates. *This means that our theory leads to a kind of granular or lattice structure of space without introducing such a strange assumption a priori".*

Max Born attempted to base a theory on momentum space and wrote a metric similar to that of Eeinstein's with the hope of introducing Heisenberg uncertainty principle and linking gravity with quantum mechanics . However that programme did not take off well even though some of key points he made are valid today as sixty years ago. Max Born did suggest that we must concede graininess at some stage.

**Tolmon** : "It is appropriate to approach the problems of cosmology with feelings of respect for their importance of awe for their vastness and of exultation for the temerity of the human mind in attempting to solve them. They must be treated, however by the detailed critical and dispassionate methods of the scientist".

**Milne** : Milne ' s ideas are particularly significant in contemplating and expanding closed universe is nearest to the model proposed in this communication. The law of gravitational attraction between two point masses is an expression in Lorentz-invariant form of the Newtonian elementary potential, with a Newtonian 'constant' of gravitation varying secularly with epoch. When converted to $\tau$-measure, it yields an elementary potential which reduces to the Newtonian form at distances not comparable with the radius of the universe, and with a 'constant' of gravitation now constant, but involving the normalization constant of the time-scale.

This formulation of the law of gravitation is capable, in the writer's opinion of accounting for the spiral character of resolved galaxies.  That the Newtonian 'constant' of gravitation varies secularly with the epoch is a     property  not confined to gravitation.  Planck's 'constant' *h* can be shown, by    later developments, to depend also secularly on the time.

**Ilya Prigoine**  : "Cosmological problems are notoriously difficult.  We still do not know what the role of gravitation was in the early universe.  Can gravitation be included in some form of the second law, or is there a kind of dialectical balance between thermodynamics and gravitation?  Certainly irreversibility could not have appeared abruptly in a time-reversible world".

**Gibbs J W**   :" The use  of the momenta instead of the velocities as independent variables is the characteristic of Hamilton's method which gives his equations of motion their remarkable degree of simplicity.  We shall find that the fundamental notions of statistical mechanics are most easily defined, and are expressed in the most simple form, when the momenta with the co-ordinates are used to describe the state of a system".

**Wheeler**  : Wheeler's  quantum foam is another attempt which comes nearest to the ideas given in this communication.  This has been recognized in the present approach by  naming  the  total  growing number of EKONS  in the universe as $N_W$.

**Hawking -  Big Bang :**  In his "area theorem' Hawking has shown that there can be "black holes"  not only in connection with big  gravitationally collapsing objects but also very small mini- micro objects  will  obey the "area theorem" and there will  be influence outside the so called black holes.  This is certainly an important result and can be reinterpreted in the light of the various physical ideas employed in the main paper.  The main difference is that in the approach of the present author it is enough to consider momentum and space part  *jointly* <u>*forming  effectively three dimensions rather than a four dimensional approach based upon the supposed need for having a   four dimensional manifold.*</u>

**Padbanabhan.T : "**Field theory works.  The miracle becomes even more curious when we notice that the bag of tricks fail miserably in the case of gravity".

**Hoyle/Narliker**  : Of the recent attempts Hoyle and Narliker proposing a quasi steady state is also significant and has some parts overlapping with the present approach.
The Hoyle  & Narliker approach   proposes that  little mini black holes are continuously taking place and that very minute iron particles  floating around are able to thermolise  the background radiation.  However these approaches are still dependent upon having time as a additional dimension and constrained by the Lorentz transformation.  They also rely upon some suitable boundary conditions in order to take care of the infinities.

**Penrose. R** :
Has been consistently advocating the Planck mass is substantially high and almost to macroscopic dimensions and tried to draw some major conclusions. However all these approaches are still rooted under the four dimensional space time. All these and many other approaches are being based on higher dimensional spaces like ten and above and innumerable theory of everything under Mbranes, strings their vibrations etc.are being proposed.

**J.G. Taylor**
"Time and space shown to have different properties, both from each other and from that of the macroscopic world, when they are investigated at distances of the size of elementary particles even the definition of time is in some doubt in such a case. Space reflection is violated maximally, but time reversal is not. Indeed it may not even be violated at all, but if it is not then causality is violated over such short distances. In any case the basic laws of physics will have to be changed; when the details are finally worked out the concept of time will be altered ineradicable from the common sense one".

# Implications of the Finite Model

## Some thoughts on possibility of understanding primary cosmic rays

If, as postulated the universe is a finite one, questions like could there not be other universes in different stages of development with their own scales of physical features, and finite dimensions and whether it is possible that there can be a clash of such cosmological entities with our own universe arise . The emphatic answer is yes, there is a possibility. May be we are already witnessing it the phenomenon of cosmic rays which are impinging with enormous energy. Possibly the primary particles of the observed cosmic rays are due to such parts of other regions impinging on ours . This possibility remains to be worked out and the consequences fitted in with available information from cosmic rays. At least the following known observations are in favour of pursuing this line of thought i) The scale of energy of the observed primary cosmic rays ii) Their uniform occurrence with reference to our cosmos.

Of course these are for detailed mathematical modeling , order of magnitude calculations and specific predictions.

**Magnetic Monopoles :** Electricity and magnetism are so deeply interconnected that the urge to find a magnetic monopole similar to the electric charge. Perhaps such an entity is to be found in a momentum-dominated universe as remarked below.

**Momentum dominated universe:** The back bone of the entire arguments adduced is the importance of a joint momentum/space approach. The universe as we know looks predominantly spatial. There is no reason why there cannot be another universe predominantly momentum dominated.

It is quite possible that such a situation may have arisen. Already with ultra high energy accelerators we may be creating such entities. Such situations may also be caused by high intensity laser beams.

## Remarks on "Theory of Everything" – M-Brane etc

Of late there has been considerable activity in extending some of these ideas in the form of strings and M-branes and other objects requiring even many more dimensions. There are many aspects of Hawkings/ Beckenstein/ Klein approaches which can be reinterpreted in the light of the new approach. Detailed reconciliation of the existing super structure already built up by eminent mathematicians with the present approach would form subject matter of future communication by the author. In this endeavor the author would appreciate constructive criticism from eminent mathematicians .
The author sees the need for the involvement of mathematicians particularly those working on large prime numbers

## Conclusion : What does this model achieve

The question raised in the title has been answered. It has been shown that Time needs to be understood as a parameter rather than as an extra dimension.
1)Finite numbers      <u>Avoid zero and infinity</u>!
2)Conceptually simple
3)Unidirection of the time concept/relation to thermo dynamics
4) Role of repulsive interaction/their origin.   Expansion of the universe follows naturally
5)Mach's principle reaffirmed and the identity of inertial and gravitational mass enabled
6)Time reversal and gauge invariance clarified naturally
7)Clarifies the relationship between the time and energy concepts
8)Clarity on fundamental constants of nature and estimation of *G* from CMBR data and Hubble's parameter
9)Clarity on the role of velocity of light and super- luminal propagation
10)Possibility of understanding high energy primary cosmic rays
11)Understanding duality through mathematics
12)New mathematics with applications reaching well beyond physics

## List of References

**Box 1**

**REMARKS BY DIRAC** a)<u>HOW TO DO THE THEORETICAL WORK :</u> ONE SHOULD NOT TRY TO ACCOMPLISH TOO MUCH IN ONE STAGE. ONE SHOULD SEPARATE THE DIFFICULTIES IN PHYSICS ONE FROM ANOTHER AS FAR AS POSSIBLE, AND THEN DISPOSE OF THEM ONE BY ONE

b)THERE ARE TWO MAIN PROCEDURES FOR A THEORETICAL PHYSICIST. ONE OF THEM IS TO WORK FROM THE <u>EXPERIMENTAL BASIS</u>. THE OTHER PROCEDURE IS TO <u>WORK FROM THE</u> MATHEMATICAL BASIS….. WITHOUT DESTROYING THE VERY GREAT SUCCESSES OF THE EXISTING THEORY !

c)<u>ON THE WRONG TRACK</u> : WE MUST REALIZE THAT THERE IS SOMETHING RADICALLY WRONG WHEN WE HAVE TO DISCARD INFINITIES FROM OUR EQUATIONS.

d)<u>COSMOLOGICAL SPECULATION</u> : IT IS USUALLY ASSUMED THAT THE LAWS OF NATURE HAVE ALWAYS BEEN THE SAME AS THEY ARE NOW. HERE IS NO JUSTIFICATION FOR THIS. THE LAWS MAY BE CHANGING, AND IN PARTICULAR QUANTITIES WHICH ARE CONSIDERED TO BE CONSTANTS OF NATURE MAY BE VARYING WITH COSMOLOGICAL TIME. SUCH VARIATIONS WOULD COMPLETELY UPSET THE MODEL MAKERS.

**REMARKS BY A.S. EDDINGTON** : THE LAW THAT ENTROPY ALWAYS INCREASES - THE SECOND LAW OF THERMODYNAMICS - HOLDS, I THINK, THE SUPREME POSITION AMONG THE LAWS OF NATURE. IF SOMEONE POINTS OUT TO YOU THAT YOUR PET THEORY OF THE UNIVERSE IS IN DISAGREEMENT WITH MAXWELL'S EQUATIONS - THEN SO MUCH THE WORSE FOR MAXWELL'S EQUATIONS. IF IT IS FOUND TO BE CONTRADICTED BY OBSERVATION - WELL, THESE EXPERIMENTALISTS DO BUNGLE THINGS SOMETIMES. BUT IF YOUR THEORY IS FOUND TO BE AGAINST THE SECOND LAW OF THERMODYNAMICS I CAN GIVE YOU NO HOPE ; THERE IS NOTHING FOR IT BUT TO COLLAPSE IN DEEPEST HUMILIATION.

Box 2

## HEISENBERG'S UNCERTAINTY RELATION REVISIED
$\Delta p_x \, \Delta \underline{x} \approx (1/2) \, \mathbf{h}$   NOT CORRECT
REALLY $(\Delta Px)$  SHOULD BE AN AREA $\perp^r$ TO $\boldsymbol{x}$
DIRECTED AREA.  MOMENT OF MOMENTUM

NOTE DIFFRACTION AT A SLIT
DEPENDS UPON THE SLIT WIDTH !
WAVE LENGTH IS ALONG DIRECTION
OF PROPAGATION

IN THE NEW FORMULTION A LESS
CONSTRAINED VERSION IS SUFFICIENT

A VOLUME OF $[½ \, (h/2\pi)]^3$
TAKEN AS THE MOST ELEMENTARY
ENTITIES CALLED <u>EKONS</u> AND
THEIR ASSEMBLY CALLED <u>CHALACHALA</u>

| Box 3 | LARGE SCALES. TIME AS A COORDINATE<br>GUIDING PRINCIPLE : EQUIVALANCE<br>GRAVITATION : FOUR DIMENSIONAL<br>REIMANIAN GEOMETRY WITH MATTER<br>ENERGY, TENSOR |

**r , t**
(r, t) 'SPACE'

---

**P, E**
( r, p, t ) SPACE
   r, p NON
COMMUTATIVE

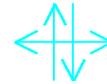 MISMATCH

GUIDING PRINCIPLE : HEISENBERG'S
UNCERTAINTY RELATIONS. SMALL LENGTH &
TIME SCALES.
MATHEMATICS IN TERMS OF MATRICES OR
DIFFERENTIAL OPERATORS. NON
COMMUTING
TIME ESSENTIALLY AS A PARAMETER

**Box 4**

**LINK : EINSTEIN'S EQUIVALENCE PRINCIPLE. POSITION & MOMENTUM LINKED TOGETHER COARSE GRAINED HAUSEDROFS SPACE.**
**LARGE TIME SCALES**
**INERTIA / EFFECTIVE MASS DUE TO STRONG MACH'S PRINCIPLE INVOKED LARGE NUMBER OF EKONS KNOTTING TOGETHER CONSTITUTE BARYONIC MATTER AT CMBR TEMPERATURE AND NON BARYONIC COLD MATTER IS TREATED AS HIGHLY DEGENERATE FERMI DIRAC DISTRIBUTION WITH ITS OWN TEMPERATURE.**

**CHALACHALA : ASSEMBLY OF FINITE NUMBER OF EKONS : ORIENTED HALF TWISTS. IDENTIFIED WITH PLANCK'S CONSTANT :**
$$[\,h/2\,]$$

**HEISENBERG'S UNCERTAINTY PRINCIPLE: JUST BEING LEFT OR RIGHT TWIST**

**TIME AS A PARAMETER**

**ONLY THREE DIMENSIONS**

**REPULSION BETWEEN LIKE TWISTS AND CANCELLATION OF OPPOSITE HALF TWISTS**

**LINK TO QUANTUM MECHANICS**
**SCHRODINGER EQUATION**
**HEISENBERG'S/DIRAC QUANTUM CONDITION**
**VERY SHORT TIME SCALE AND HIGH ENERGY**

# JUMPING BOY & THE EARTH KICKS THE EARTH

I KNOW EINSTEIN !

HI ! I HAVE KICKED THE EARTH, I HAVE CREATED
$½ M_{EARTH} \times V^2$

BOY JUMPS

EARTH SMILES YOU LITTLE BRAT YOU HAVE ONLY
$½ m_{BOY} \times V^2$
(ENERGY)

WHO IS RIGHT ?

Box 5

# GRAPHICAL REPRESENTATION OF A CLOCK

## Archemedian Spiral: Polar Plot

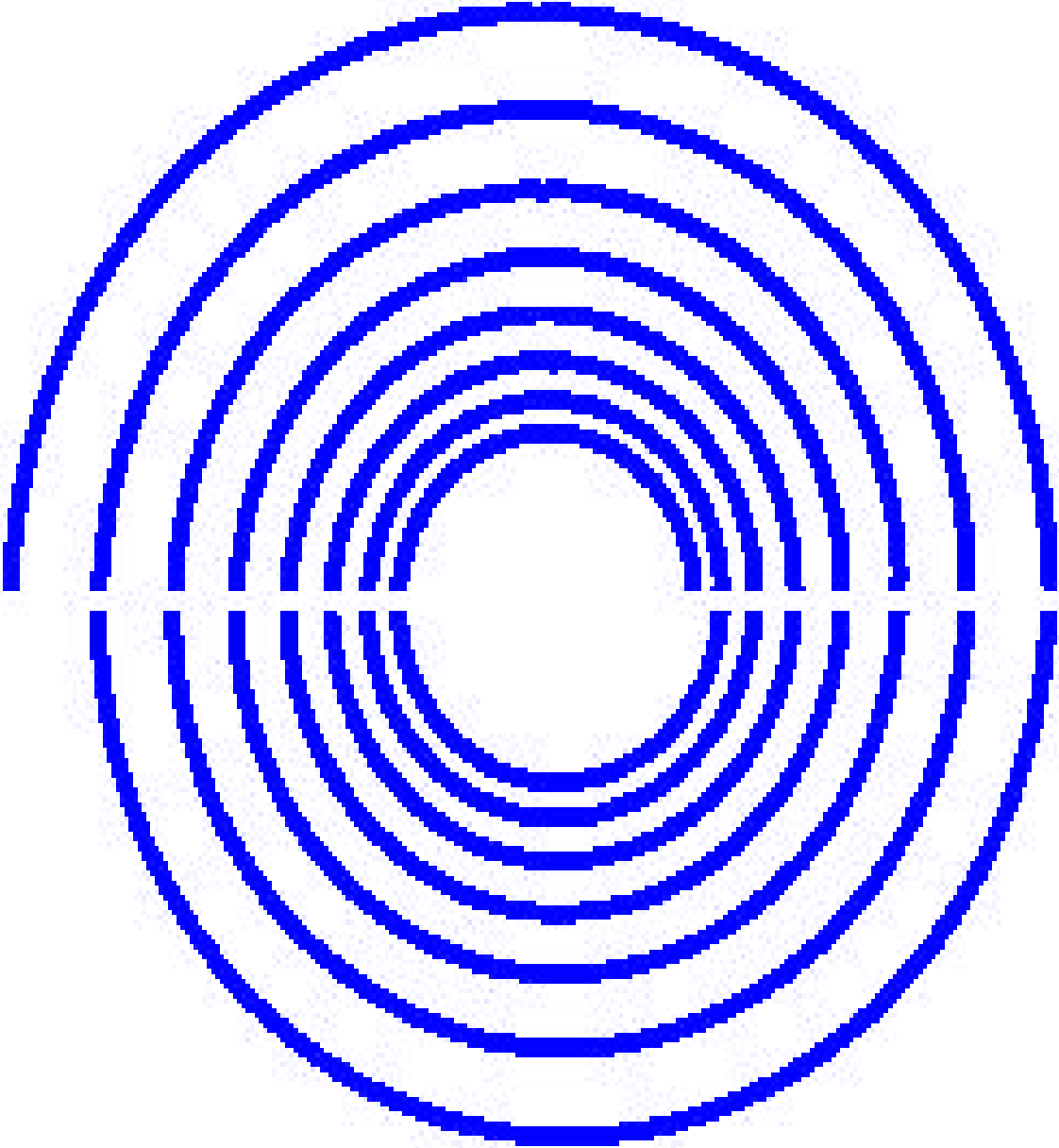

Box 6

BOX 7

DUALITY BETWEEN HOMOLOGY (CONTINUOUS)
AND COHOMOLOGY (DISCRETE)

HOMOLOGICAL OPERATIONS MEASURE THE NATURE
OF THE CONTINUITY STRUCTURE OF THE MANIFOLD

BY PERFORMING CONTINUOUS OPERATIONS
(THE SO CALLED CONTRACTION) ON CLOSED CURVES GOING
THROUGH THE POINTS OF THE MANIFOLD.

"DUALITY" WE HAVE ..TO POINT OUT IS A
FUNDAMENTAL ONE IN MANY MATHEMATICAL
DISCIPLINES ; REVEALING SOMETHING NON TRIVIAL ABOUT THEM

STANDARD MATHEMATICAL THEORY, SET THEORY LOGIC

| (A) DISCRETE (ALGEBRA) | (B) CONTINUOUS (TOPOLOGY) |
|---|---|
| GROUP, RINGS GRAPHS, ETC. | TOPOLOGICAL GROUPS RINGS ETC |
| ALGEBRAIC GEOMETRY | ANALYTIC GEOMETRY |
| ALGEBRAIC TOPOLOGY | ANALYTIC TOPOLOGY |
| (HOMOLOGICAL ALGEBRA) | MANIFOLDS |
| ALGEBRAIC LIE THEORY | LIE THEORY |
| ALGEBRAIC SHEAF THEORY | ANALYTIC SHEAF THEORY |
| ALGEBRAIC NUMBER THEORY | ANALYTIC NUMBER THEORY |
| ALGEBRAIC MANIFOLDS | ANALYTIC (& DIFFERENTIABLE) MANIFOLDS |
| COMBINATORIAL THEORY OF PROBABILITY | ANALYTIC THEORY OF PROBABILITY (MEASURE THEORY) |
| SUMMABILITY THEORITIES (SERIES) | INTEGRAL CALCULUS |
| DIFFERENCE EQUATIONS | DIFFERENTIAL EQUATIONS |
| LINEAR ALGEBRA | LINEAR ANALYSIS (BANACH ALGEBRA'S) |
| FINITE GEOMETRIES (COMPUTER SCIENCE) | CLASSIC GEOMETRIES |

Category consists of all objects of a certain sort groups, topological spaces etc. and of the mappings (homomorphisms) continuous maps etc. that exists between the objects.

Willen Kyuk Complementarity in Mathematics and its application.
D.Riedel Publication Limited 1977